\documentclass{nature}

\usepackage{graphicx}
\usepackage{lineno}
\usepackage{mathtools}
\title{Chiral Spin-Liquid-Like State in Pyrochlore Iridate Thin Films}

\author{Xiaoran Liu$^{1,2,\ast}$, Jong-Woo Kim$^{3}$, Yao Wang$^{4}$,  Michael Terilli$^{2}$, Xun Jia$^{5,6}$, Mikhail Kareev$^{2}$, Shiyu Peng$^{1,7}$, Fangdi Wen$^{2}$, Tsung-Chi Wu$^{2}$, Huyongqing Chen$^{8}$, Wanzheng Hu$^{8,9,10}$, Mary H. Upton$^{3}$, Jungho Kim$^{3}$, Yongseong Choi$^{3}$, Daniel Haskel$^{3}$, Hongming Weng$^{1,7}$, Philip J. Ryan$^{3}$, Yue Cao$^{6}$, Yang Qi$^{4}$, Jiandong Guo$^{1,7}$ \& Jak Chakhalian$^{2}$}

\begin{document}

\maketitle

\begin{affiliations}
 \item Beijing National Laboratory for Condensed Matter Physics and Institute of Physics, Chinese Academy of Sciences, Beijing 100190, China.
 \item Department of Physics and Astronomy, Rutgers University, Piscataway, New Jersey 08854, USA.
 \item X-ray Science Division, Argonne National Laboratory, Lemont, Illinois 60439, USA.
 \item State Key Laboratory of Surface Physics and Department of Physics, Fudan University, Shanghai 200433, China.
 \item Multi-disciplinary Research Division, Institute of High Energy Physics, Chinese Academy of Sciences, Beijing 100049, China.
 \item Materials Science Division, Argonne National Laboratory, Lemont, Illinois 60439, USA.
 \item School of Physical Sciences, University of Chinese Academy of Sciences, Beijing 100049, China.
 \item Department of Physics, Boston University, Boston, Massachusetts 02215, USA.
 \item Division of Materials Science and Engineering, Boston University, Boston, Massachusetts 02215, USA.
 \item Photonics Center, Boston University, Boston, Massachusetts 02215, USA.

 \end{affiliations}

\begin{abstract}

The pyrochlore iridates have become ideal platforms to unravel fascinating correlated and topological phenomena that  stem from the intricate interplay among strong spin-orbit coupling, electronic correlations, lattice with geometric frustration, and itinerancy of the 5\textit{d} electrons. 
The all-in-all-out antiferromagnetic state, commonly considered as the magnetic ground state, can be dramatically altered in reduced dimensionality, leading to exotic or hidden quantum states inaccessible in bulk. Here, by means of magnetotransport, resonant elastic and inelastic x-ray scattering experiments, we discover an emergent quantum disordered state in (111) Y$_2$Ir$_2$O$_7$ thin films (thickness $\leq$30 nm) persisting down to 5 K, characterized by dispersionless magnetic excitations. The anomalous Hall effect observed below an onset temperature near 135 K corroborates the presence of chiral short-range spin configurations expressed in non-zero scalar spin chirality, breaking the macroscopic time-reversal symmetry. 
The origin of this chiral state is ascribed to the restoration of magnetic frustration on the pyrochlore lattice in lower dimensionality, where the competing exchange interactions together with enhanced quantum fluctuations suppress any long-range order and trigger spin-liquid-like behavior with degenerate ground-state manifold. 
\end{abstract}

\newpage

%{\noindent \bf Introduction}\\
In the past decade, the field of emergent many-body phenomena in solids has undergone a paradigm shift driven by the convergence of two major streams of quantum materials: band topology and electron-electron interaction \cite{Tokura_NP_2017}. In heavy transition metal compounds with comparable strength of spin-orbit coupling (SOC) and on-site Coulomb repulsion, the electron band topology is intertwined with magnetism, leading to novel correlated topological phases with spontaneous time-reversal symmetry breaking \cite{Balents_ARCMP_2014, nenno_nrp_2020, smejkal_nrm_2022}. 
Pyrochlore iridates $R_2$Ir$_2$O$_7$ ($R$ = Y and rare-earth elements) are a representative model system that has received considerable attention for exploring the delicate balance between SOC, correlations, and the itinerancy of 5{\it d} electrons embedded on the geometrically frustrated pyrochlore lattice. From theoretical viewpoint, extremely rich phase diagrams of $R_2$Ir$_2$O$_7$ have been constructed with a plethora of interesting correlated topological phenomena \cite{Balents_NatPhys_2010,Wan_PRB_2011,WK_PRB_2012,Wang_PRB_2017,Go_PRL_2012,Varnava_PRB_2018,Wang_PRL_2017,Ladovrechis_PRB_2021,Shinaoka_PRL_2015,Savary_PRX_2014,Hwang_PRL_2020,Yamaji_PRX_2014}. 
Recently, the link between topological properties and dimensionality has further inspired the predictions and search for emergent hidden quantum states that are only accessible in (111)-oriented thin films of pyrochlore iridates \cite{Hu_PRB_2012, Nagaosa_PRL_2014, Hwang_SR_2016, Chen_2015_PRB, Fiete_PRL_2017,Ohtsuki_PNAS_2019,Kim_SA_2020,Li_AM_2021,Liu_PRL_2021}.

\begin{figure}[htp]
\centering
\includegraphics[width=0.8\textwidth]{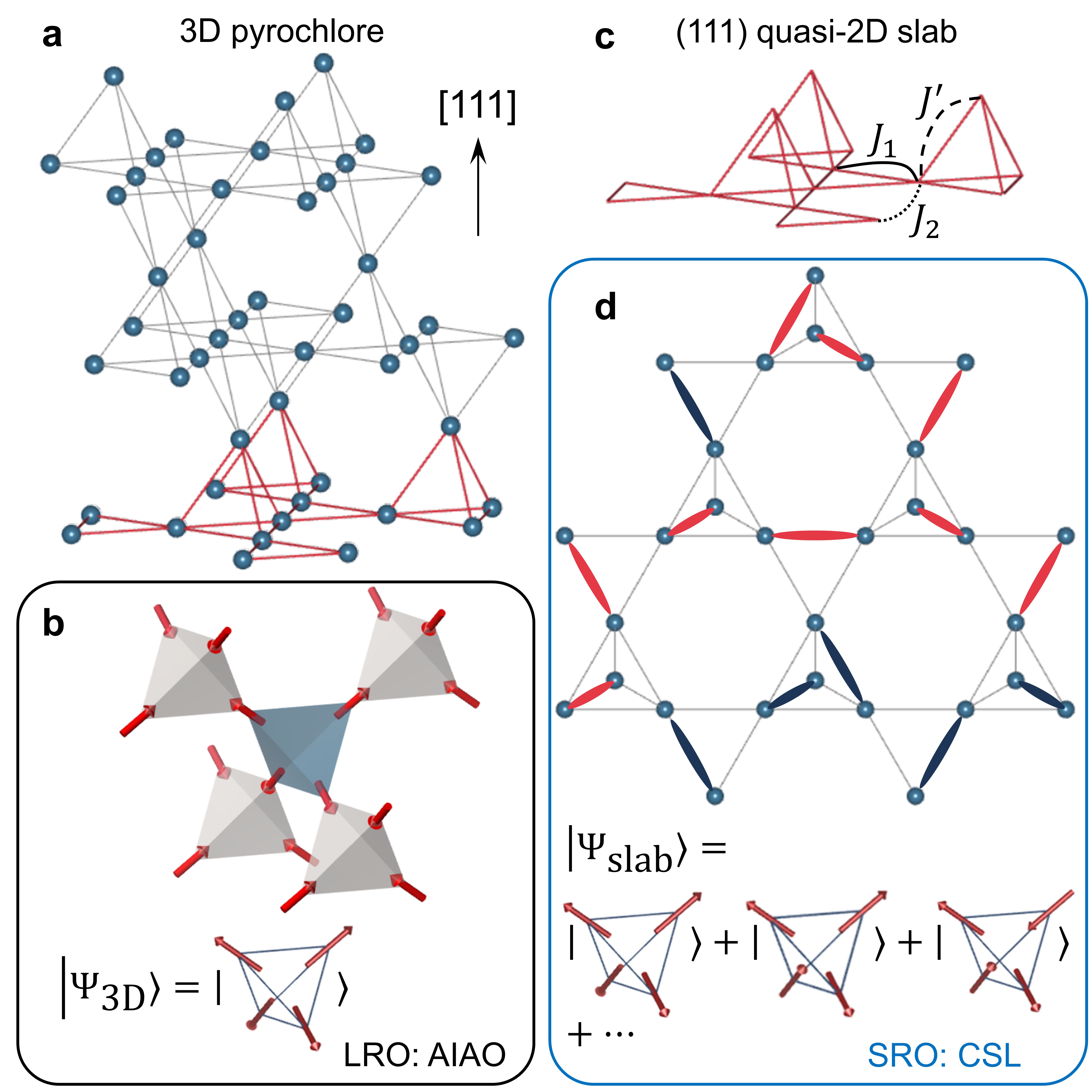}
\caption{\label{structure} {\bf Schematics of exotic magnetism on (111) Y$_2$Ir$_2$O$_7$ lattice.} 
\textbf{a}, 3D pyrochlore framework constructed by Ir atoms. 
\textbf{b}, Long-range ordering with the AIAO antiferromagnetic spin configuration on Ir sublattice. 
\textbf{c}, (111) quasi-2D slab of pyrochlore lattice composed of alternating kagome and triangle atomic layers. $J_1$ ($J_2$) represents the Ir nearest-neighbor (next-nearest-neighbor) interaction within the kagome layers, while $J^{\prime}$ the nearest-neighbor interaction between kagome and triangle layers.   
\textbf{d}, A possible snapshot of the chiral spin liquid configuration on (111) quasi-2D pyrochlore slab. The Ir spins pair up forming dimers covering the entire lattice. The two spin directions in one dimer can be either the same (red dimer: both towards or away from the center of tetrahedra) or opposite (blue dimer: one towards and one away from the center of tetrahedra).}
\end{figure}

In bulk, the high temperature `parent' phase of $R_2$Ir$_2$O$_7$ is the Luttinger-Abrikosov-Beneslavskii (LAB) non-Fermi-liquid semimetal, with a quadratic band touching point located at the zone center \cite{Moon_PRL_2013,Kondo_NC_2015,Nakayama_PRL_2016}. At low temperatures, formation of magnetic long-range ordering (LRO) on the Ir sublattice can turn the LAB semimetal into various topological Weyl states \cite{Moon_PRL_2013}. 
%The presence of Weyl nodes in PyIr is the fundamental source of several fascinating magneto-transport phenomena, including the anomalous Hall effect (AHE) in a zero external magnetic field, saturated longitudinal resistivity at low temperatures, chiral anomaly, and the planar Hall effect[REFS]. 
%
The magnetic interactions between Ir$^{4+}$ ions on the corner-sharing tetrahedral network are governed by two leading terms: antiferromagnetic (AFM) Heisenberg exchange and Dzyaloshinskii-Moriya (DM) interaction. The AFM interaction can induce strong frustration and lead to a highly degenerate spin-liquid ground state. In contrast, the DM interaction tends to lift the degeneracy and stabilize LRO with the peculiar all-in-all-out (AIAO) spin configuration  \cite{Elhajal_PRB_2005,sagayama_PRB_2013,donnerer_PRL_2016,Tomiyasu_JPSJ_2012,Lefrancois_PRL_2015,Disseler_PRB_2014,Lefrancois_NC_2017} [Fig.~\ref{structure}b]. Furthermore, the inclusion of pseudo-dipolar interaction, single-ion anisotropy, or the variation of rare-earth $R$ ion rapidly expands the phase space, stabilizing twelve symmetry-allowed $\textbf{q}$ = 0 spin configurations and their linear combinations as the plausible magnetic ground state. \cite{Wang_PRL_2017,Ladovrechis_PRB_2021,WK_PRB_2012,Chen_PRR_2020}

In the quasi-two-dimensional (quasi-2D) limit, however, the tetrahedral network can be transformed into a thin (111)-oriented slab composed of several kagome and triangle atomic layers. In this case, one can envision how the competition among multiple AFM exchanges may restore strong frustration and stabilize an exotic short-range order (SRO) with finite scalar spin chirality \cite{kawamura_prl_2002}. Conceptually, this novel magnetic state is akin to a chiral spin liquid (CSL), where fluctuating spin dimers cover the entire pyrochlore slab and break the global time-reversal symmetry \cite{wen_prb_1989,messio_prl_2017}. As such, it is interesting to explore the challenging question of the true magnetic ground state of pyrochlore iridates on the thin (111)-oriented slab. 

For this purpose, we have developed a set of high-quality (111) Y$_2$Ir$_2$O$_7$ films using the {\it in-situ} solid phase epitaxy method \cite{Liu_APL_2019}. By means of resonant x-ray magnetic scattering techniques and magneto-transport measurements, we have discovered the formation of $\textbf{q}$ = 0 AFM LRO in 100 nm (111) Y$_2$Ir$_2$O$_7$ films at a N\'{e}el temperature near 140 K. Surprisingly, the LRO is absent in thin films ($\leq$30 nm). In this quasi-2D limit, a novel CSL state with anomalous Hall effect (AHE) appears below a characteristic temperature around 135 K, stemming from the fluctuating non-coplanar spin textures with non-zero chirality. In sharp contrast to the dispersive magnon mode observed in bulk crystals and predicted from the linear spin-wave theory, the CSL state exhibits spin-gapped but dispersionless spectrum of magnetic excitations. Our findings highlight the previously unexplored role of dimensional confinement to stabilize emergent quantum states in thin slabs of pyrochlore iridates for the first time.\\

\begin{figure}[htp]
\centering
\includegraphics[width=0.9\textwidth]{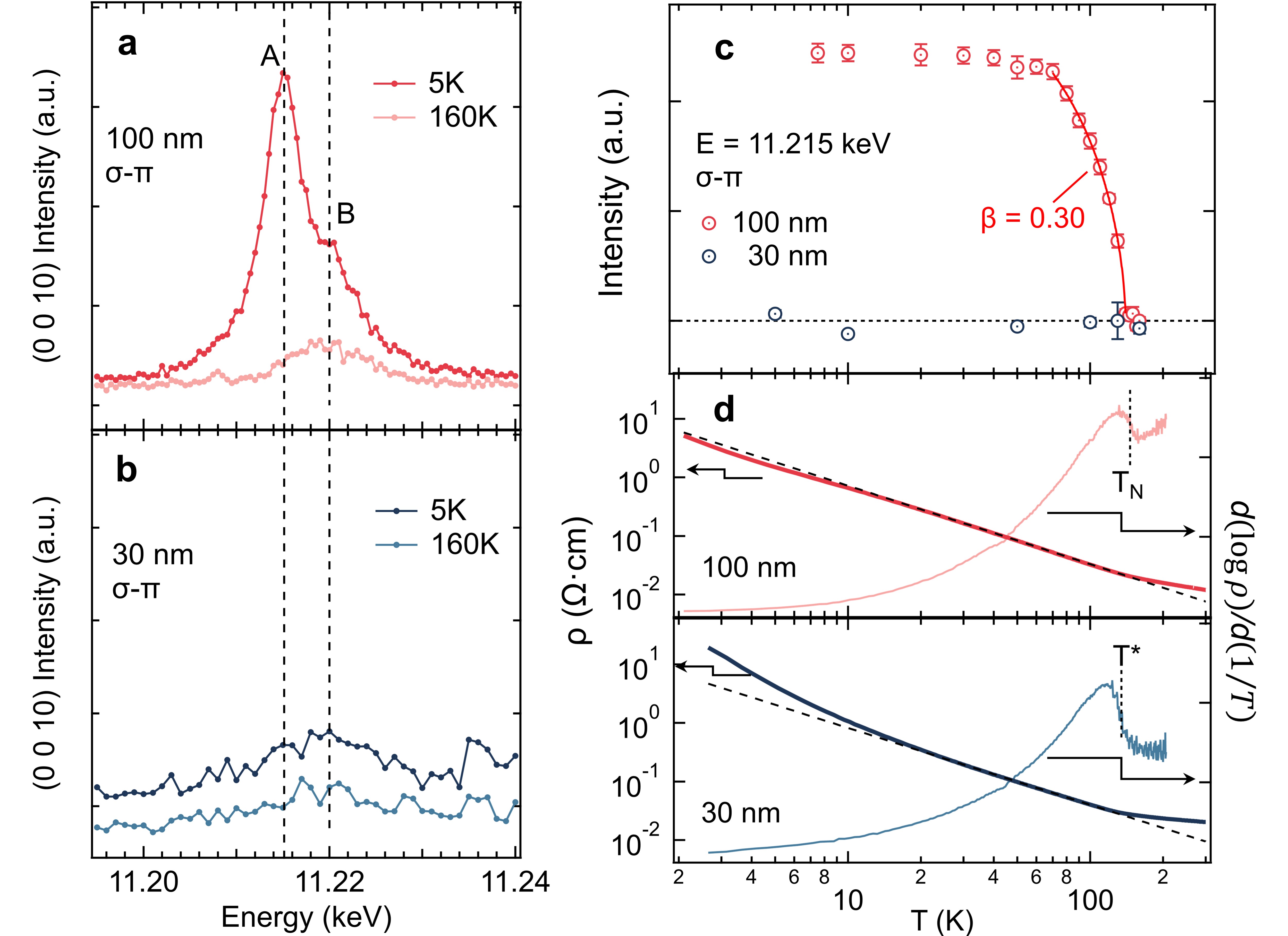}
\caption{\label{RMS} {\bf Magnetic phase transition in 100 nm and 30 nm (111) Y$_2$Ir$_2$O$_7$ films.}  
\textbf{a}-\textbf{b}, Resonance profiles of the magnetic (0 0 10) reflection near Ir L$_3$ edge in the $\sigma$-$\pi$ channel for (\textbf{a}) 100 nm and (\textbf{b}) 30 nm Y$_2$Ir$_2$O$_7$ films. A special azimuthal alignment which includes setting the [110] axis of the Y$_2$Ir$_2$O$_7$ films perpendicular to the scattering plane allows for effectively suppressing the ATS contribution and leaving the RMS signal dominant in the $\sigma$-$\pi$ channel \cite{sagayama_PRB_2013}. Peak A at 11.215 keV represents the RMS signal, which is significantly enhanced at 5 K in the long-range antiferromagnetic state. Peak B at 11.22 keV is due to the leakage of ATS that is dominant in the $\sigma$-$\sigma$ channel.   
\textbf{c}, Temperature dependence of the integrated intensity of magnetic (0 0 10) reflection for 100 nm (red) and 30 nm (blue) Y$_2$Ir$_2$O$_7$ films. The transition temperature $T_N$ $\approx$ 140 K and the critical exponent $\beta$ $\approx$ 0.30 of 100 nm sample are extracted from the data following the relationship $I \propto M^2 \propto (1-\frac{T}{T_N})^{2\beta}$.
\textbf{d}, Temperature dependence of the resistivity of 100 nm (red) and 30 nm (blue) samples, which exhibit a ``kink'' behavior at around 145 K and 135 K, respectively. The black dashed lines are power-law fits ($\rho \propto T^{-\alpha}$) of the data between 100 -- 20 K, with the power index $\alpha$ $\sim$1.3 (1.4) for 100 nm (30 nm) sample.}
\end{figure}

{\noindent \bf Resonant X-ray diffraction}\\
It is generally recognized that upon lowering the temperature, all $R_2$Ir$_2$O$_7$ except for $R$ = Pr undergo a metal-to-insulator/semimetal transition concurrently with the AFM phase transition \cite{Gardner_RMP_2010}. In particular, since bulk Y$_2$Ir$_2$O$_7$ shows the highest N\'{e}el temperature $\sim$150 K and has no $f$-electron moments on the Y sublattice, it serves as an ideal playground to explore the intrinsic magnetism of the Ir$^{4+}$ sublattice. Despite active experimental efforts, inconsistent results on the nature of the ground state have been reported. For instance, well-defined oscillations of single frequency were observed in muon spin resonance/relaxation spectra, indicating the formation of LRO \cite{Disseler_PRB_2012}. But meanwhile, neutron powder diffraction data yielded no sign of magnetic peaks at low temperatures, attributed to high neutron absorption of Ir \cite{Shapiro_PRB_2012}. With the availability of high-quality single-crystalline films, we use the resonant X-ray diffraction (RXD) at the Ir $L_3$ edge to address this fundamental question. 

The AIAO spin structure is a $\textbf{q}$ = 0 LRO, which triggers charge-forbidden reflections such as (0 0 4n+2) visible in the resonant magnetic scattering (RMS). However, due to the local trigonal distortion of the IrO$_6$ octahedra, the anisotropic tensor susceptibility (ATS) scattering is also attributed to the (0 0 4n+2) reflections \cite{dmitrienko_AC_2005}. 
As seen in Fig.~\ref{RMS}a, in 100 nm Y$_2$Ir$_2$O$_7$, the magnetic Bragg signal is clearly observed at 5 K in the energy spectra of (0 0 10) reflection at peak A ($E$ = 11.215 keV); the signal gradually diminishes to zero at 160 K. The hump feature at peak B ($E$ = 11.22 keV) is a leakage of the ATS scattering into the $\sigma$-$\pi$ channel, and its intensity barely varies with temperature (Supplementary Fig.~S2). 
The temperature dependence of the (0 0 10) integrated intensity indicates the onset of magnetic phase transition near 140 K in 100 nm Y$_2$Ir$_2$O$_7$ [Fig.~\ref{RMS}c]. These observations are in good agreement with the reported bulk value, demonstrating the establishment of LRO on the Ir sublattice \cite{Disseler_PRB_2012}. A power-law fit of the scattered intensities in the vicinity of transition transition, $I \propto M^2 \propto (1-\frac{T}{T_N})^{2\beta}$ yields the critical exponent $\beta \approx$ 0.30, falling into the 3D Ising universality class consistent with the AIAO spin configuration \cite{Savary_PRX_2014}.

In stark contrast, direct examination of Fig.~\ref{RMS}b immediately reveals that, other than the small leakage of ATS in the $\sigma$-$\pi$ scattering channel, the RMS signal is no longer detectable in 30 nm Y$_2$Ir$_2$O$_7$ in a wide range of temperatures down to 5 K! Moreover,  the absence of the expected magnetic scattering signal is verified on the set of (0 0 4n+2) reflections. A potential sample quality issue is ruled out, as the charge scattering peaks allowed for the pyrochlore lattice and the presence of characteristic ATS contributions are all detected [see Supplementary Fig.~S1 and Fig.~S2 for more details]. 
On the other hand, the X-ray magnetic circular dichroism (XMCD) spectra at 4 K have shown that the averaged net magnetic moment per Ir$^{4+}$ ion is approximately 0.023(4) $\mu_\textrm{B}$/Ir [Supplementary Fig.~S3]. This is remarkably lower than the expected local moment of about 0.33 $\mu_\textrm{B}$/Ir for the $J_\textrm{eff} = 1/2$ configuration \cite{Shapiro_PRB_2012}, indicating an unusual paramagnetic (PM) behavior in 30 nm Y$_2$Ir$_2$O$_7$ even under 6 T magnetic field, where the local moments of Ir persistently fluctuate at low temperatures.\\

%A rough estimate from the high-temperature data yields an activation gap of Y$_2$Ir$_2$O$_7$ less than $\sim$15 meV. The ``$d(ln\rho)/d(1/T)$ vs. T'' plot [right panel of Fig.~\ref{RMS}d] indicates the possible existence of magnetic transition at $\sim$140 K (135 K) in 100 nm (30 nm) Y$_2$Ir$_2$O$_7$. 
%Notably, a distinguishable metal-semimetal transition is captured on 100 nm Eu$_2$Ir$_2$O$_7$ with T$_\textrm{c}$ $\sim$105 K. Since both Y$^{3+}$ and Eu$^{3+}$ ions are non-magnetic, their impact on the transport behaviors likely manifest the differences in the underlying topological and magnetic properties, due to variation the hopping amplitudes of Ir electrons \cite{Ueda_PRB_2016}.\\ 

\begin{figure}[htp]
\centering
\includegraphics[width=0.8\textwidth]{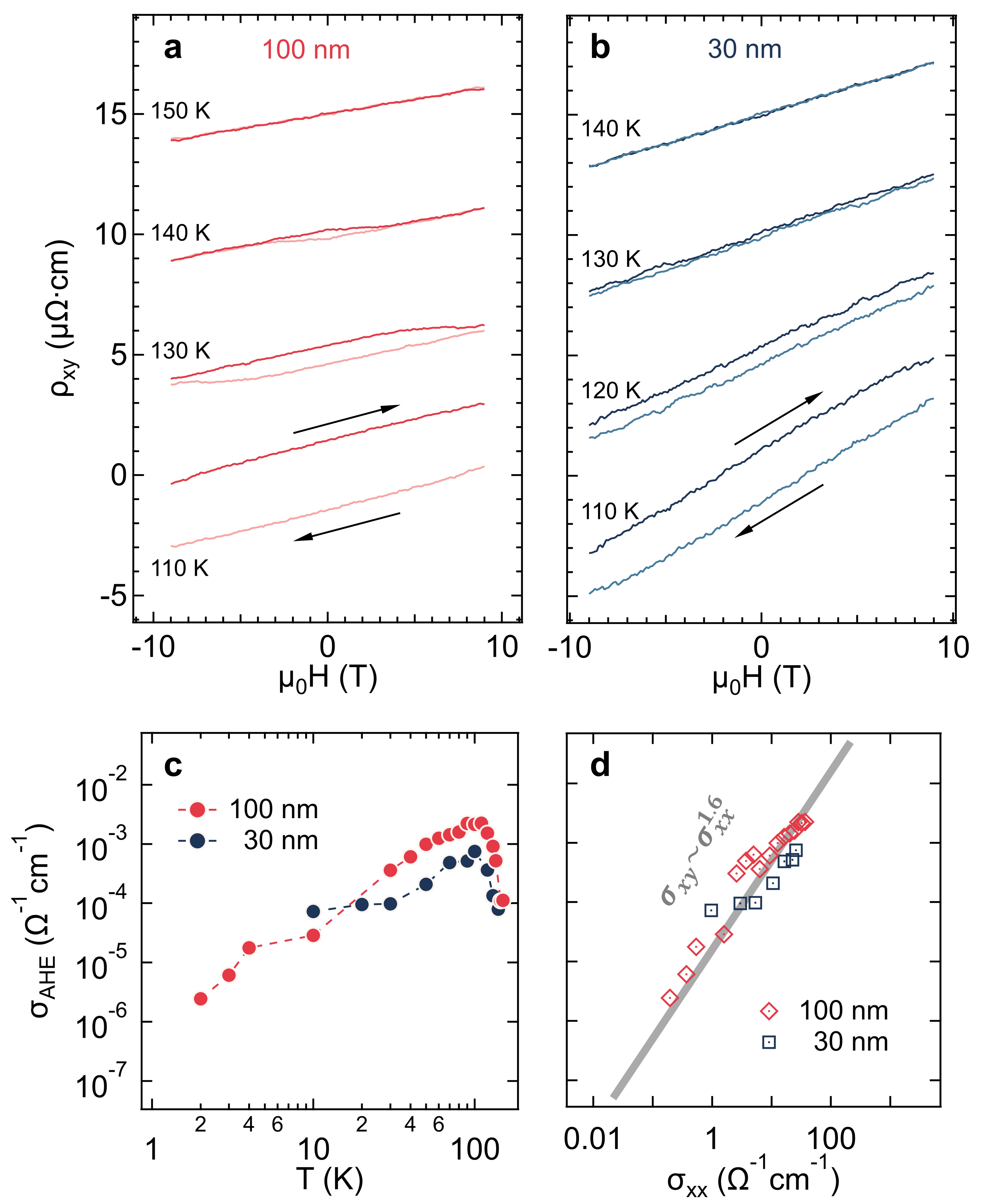}
\caption{\label{AHE} {\bf Anomalous Hall effect.} 
\textbf{a}-\textbf{b}, Evolution of the transverse magneto-resistivity $\rho_{xy}$ measured at a set of temperatures across (\textbf{a}) $T_N$ of 100 nm and (\textbf{b}) $T^*$ of 30 nm sample. Each branch of the curve (in light or heavy color) is obtained after field cooling, with an black arrow showing the corresponding scan direction. For clarity, the center of curve at each temperature is offset by 5 $\mu\Omega\cdot$cm.
\textbf{c}, Temperature dependence of the spontaneous Hall conductivity $\sigma_\textrm{AHE}$ of 100 nm and 30 nm Y$_2$Ir$_2$O$_7$ films. 
\textbf{d}, The scaling relationship between $\sigma_\textrm{AHE}$ and $\sigma_{xx}$ of Y$_2$Ir$_2$O$_7$ films.}
\end{figure}

{\noindent \bf Time-reversal symmetry breaking}\\
Next, we focus on the magneto-transport response of our (111)-oriented films. In bulk Y$_2$Ir$_2$O$_7$, the PM to AFM transition manifests itself as a `kink' on the temperature-dependent resistivity curve due to the inclusion of additional scattering channels from the long-range ordered spins. 
As seen in Fig.~\ref{RMS}d, both films exhibit a semiconducting behavior with the kink feature on the $\rho$(T) curve. The kink of 100 nm Y$_2$Ir$_2$O$_7$ indeed appears near 145 K, coinciding with the N\'{e}el temperature T$_N$ obtained from our magnetic scattering data [Fig.~\ref{RMS}d, upper panel]. 
Surprisingly, the kink is also present in 30 nm Y$_2$Ir$_2$O$_7$ near T$^* \approx 135$ K [Fig.~\ref{RMS}d, bottom panel], which signals the existence of a hidden order not captured by magnetic scattering. 

The anomalous Hall effect (AHE) is a direct and macroscopic probe of time-reversal symmetry in quantum antiferromagnets \cite{smejkal_nrm_2022}. Our Hall transport data reveal the presence of distinct AHE at zero magnetic field for both Y$_2$Ir$_2$O$_7$ films, signifying that the time-reversal symmetry is spontaneously broken at low temperatures [Fig.~\ref{AHE}a-b]. Specifically, the transverse resistivity $\rho_{xy}$ of 100 nm Y$_2$Ir$_2$O$_7$ displays a clear hysteresis loop right below the N\'{e}el temperature. 
For 30 nm Y$_2$Ir$_2$O$_7$, despite the absence of LRO, non-zero spontaneous Hall resistivity starts to emerge at 130 K, coinciding with the characteristic temperature T$^*$ obtained from the longitudinal $\rho_{xx}$(T) curve. It is also worth noting that the coercive fields of both samples increase rapidly as temperature decreases, making only a small portion of the loops visible [see Methods and Supplementary Fig.~S5]. 

In addition, both films exhibit a non-monotonic temperature dependence of spontaneous Hall conductivity $\sigma_\textrm{AHE}$ [Fig.~\ref{AHE}c]. For 100 nm Y$_2$Ir$_2$O$_7$, the emergence of non-zero $\sigma_\textrm{AHE}$ concurrently with the onset of the AIAO AFM order agrees with the theoretical prediction for the formation of Weyl semimetallic state in (111) pyrochlore iridate films \cite{Hu_PRB_2012,Nagaosa_PRL_2014,Hwang_SR_2016}. Here, the incomplete cancellation of the Chern vectors over all Weyl pairs triggers an intrinsic AHE, as demonstrated recently in Eu$_2$Ir$_2$O$_7$ films \cite{Liu_PRL_2021}.
Notably, symmetry analysis based on magnetic group theory strictly enforces zero scalar spin chirality for the AIAO configuration \cite{Suzuki_PRB_2021}. The non-monotonic $\sigma_\textrm{AHE}$(T) behavior (i.e., $\sigma_\textrm{AHE}$ first increases to the maximum and gradually decreases on cooling) in 100 nm Y$_2$Ir$_2$O$_7$ thus reflects the evolution of the Berry curvatures in momentum-space, due to the motion of the Weyl nodes with decreasing temperatures \cite{Yamaji_PRX_2014}.
In contrast, because of the absence of LRO in 30 nm Y$_2$Ir$_2$O$_7$, it precludes the formation of Weyl nodes in this film. Meanwhile, the minute averaged magnetization obtained from XMCD at the Ir L-edge under 6 T excludes the origin of AHE from aligned moments in a conventional ferromagnet or a spin glass [Supplementary Fig.~S3]. These findings strongly imply the presence of a chiral spin-liquid-like state below T$^*$, with $\sigma_\textrm{AHE}$ directly measuring the scalar spin chirality averaged over all short-range non-coplanar configurations \cite{wen_prb_1989}.

Next, we explore the scaling relation between the longitudinal conductivity $\sigma_{xx}$ and the $\sigma_\textrm{AHE}$. Conventionally, this relationship can be divided into three empirical regimes based on the value of $\sigma_{xx}$ \cite{Onoda_PRB_2008, Nagaosa_RMP_2010}. As depicted in Fig.~\ref{AHE}d, both Y$_2$Ir$_2$O$_7$ films belong to the bad-metal-hopping regime, where $\sigma_\textrm{AHE}$ decreases as $\sigma_{xx}$ decreases, following the universal scaling of $\sigma_{xy} \sim \sigma_{xx}^{1.6}$ at temperatures below 100 K. This scaling behavior is generally attributed to the \textit{intrinsic} Berry-phase contribution being suppressed by scattering-dependent events from disorders with strong spin-orbit interaction \cite{Nagaosa_RMP_2010}.\\

% Note, since the Berry curvature and spin chirality interpretations are momentum- and real-space counterparts, these results provide further evidence on the intrinsic nature of AHE in Y$_2$Ir$_2$O$_7$ films.\\

\begin{figure}[htp]
\centering
\includegraphics[width=0.8\textwidth]{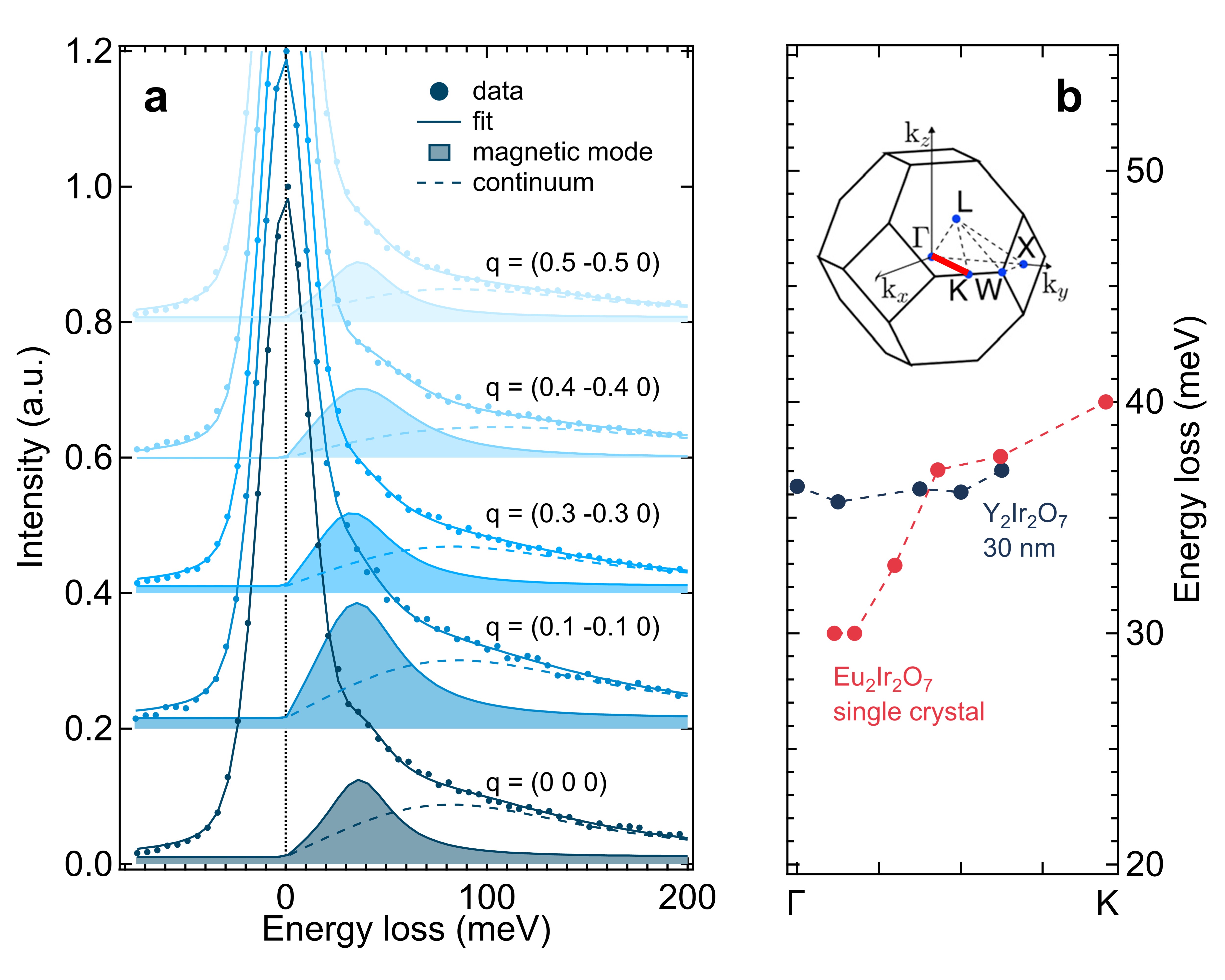}
\caption{\label{RIXS} {\bf Magnetic excitation in 30 nm Y$_2$Ir$_2$O$_7$ film.} 
\textbf{a}, The stacking RIXS spectra along the $\Gamma$-K direction recorded at 10 K. The data (solid dots) are normalized with respect to the elastic scattering peak at 0 meV. Fitting of each spectrum yields the elastic background (solid curve), inelastic peak from the magnetic excitation (shadowed curve), and inelastic background continuum (dashed curve). \textbf{b}, The extracted dispersion of magnetic excitation in 30 nm Y$_2$Ir$_2$O$_7$ film along the $\Gamma$-K direction. The inset illustrates the Brillouin zone with notations for high-symmetry positions. For comparison, the dispersive magnon reported in Eu$_2$Ir$_2$O$_7$ bulk crystal\cite{Chun_PRL_2018} with the AIAO LRO is plotted together on the figure.
}
\end{figure}

{\noindent \bf Resonant inelastic X-ray scattering}\\
In the presence of the CSL state in 30 nm Y$_2$Ir$_2$O$_7$, one can also anticipate the breakdown of classical spin-wave theory, as the spin excitations may exhibit marked deviations from the magnon modes of the bulk. In single crystals of pyrochlore iridates with the AIAO magnetic ground state,   
%Specifically, the magnon bands of R$_2$Ir$_2$O$_7$ with the AIAO order are derived from the following Hamiltonian   
%\begin{align}
%H = \sum_{ij}[J \mathbf{S_i} \cdot \mathbf{S_j} + \mathbf{D_{ij}} \cdot (\mathbf{S_i} \times \mathbf{S_j})]
%\end{align}
%where $J$ and $D$ are the AFM Heisenberg and DM interactions, respectively. 
dispersive magnon modes with a spin gap of 25 meV and 28 meV have been probed by resonant inelastic X-ray scattering (RIXS) in Sm$_2$Ir$_2$O$_7$ \cite{donnerer_PRL_2016} and Eu$_2$Ir$_2$O$_7$ \cite{Chun_PRL_2018}. 
Fig.~\ref{RIXS}a demonstrates a set of RIXS spectra taken on 30 nm Y$_2$Ir$_2$O$_7$ at 10 K, with the energy loss range below 300 meV taken along the $\Gamma$-K direction of the Brillouin zone.
The overall spectrum is composed of several distinct contributions. These include a peak at zero energy due to elastic scattering (solid curve) and two other peaks next to the elastic line (shadowed curve and dashed curve). The mode associated with the shadowed peak has an energy close to the spin gap observed in other pyrochlore iridates and is interpreted as a magnetic excitation. The dashed peak represents a particle-hole continuum that is incoherent.
%since the energy locates below 100 meV, it is unlikely for orbital excitations in any pyrochlore iridate. 

The overall dispersion of the magnetic excitation of 30 nm Y$_2$Ir$_2$O$_7$ is shown in Fig.~\ref{RIXS}b. As immediately seen, it exhibits a surprisingly dispersionless feature over a momentum range larger than half Brillouin zone along the $\Gamma$-K direction. This unusual behavior is in sharp contrast to the previously reported RIXS data on pyrochlore iridates with the AIAO LRO, where gapped and dispersive magnon modes have been observed in the energy range from 20 meV to 40 meV along this direction \cite{donnerer_PRL_2016,Chun_PRL_2018}. 
%Granted, for other types of long-range magnetic orders, dispersive magnon modes are generally expected from spin-wave theory (e.g. the XY-type AFM state hosts dispersive but gapless magnon modes \cite{donnerer_PRL_2016}). 
%In addition, our RIXS data on 100 nm Y$_2$Ir$_2$O$_7$ demonstrate the dispersive magnon mode close to bulk Eu$_2$Ir$_2$O$_7$ and reveal it can persist even in the paramagnetic phase above T$_N$ \cite{Michael_arXiv}. 
Such a dispersionless magnetic excitation observed in 30 nm Y$_2$Ir$_2$O$_7$ necessitates a fundamentally different magnetic Hamiltonian in the quasi-2D case. While the exact form is unknown, the CSL state is plausibly stabilized due to the competition among multiple exchange interactions, melting the LRO and generating chiral spin textures imposed on the triangle and kagome structural units.\\ 

{\noindent \bf Discussion}\\ 
The CSL state breaks the time-reversal symmetry, while remains magnetically disordered and entangled, fluctuating among the massively degenerate states \cite{wen_prb_1989}. For bulk pyrochlore iridates, the Ir twelve types of $\textbf{q}$ = 0 magnetic basis vectors $\psi_i$ are categorized into four irreducible representations $\Gamma_{3,5,7,9}$ (Supplementary Fig.~S6). Notably, to break the macroscopic time-reversal symmetry, only those six bases under $\Gamma_9$ are allowed to generate nonvanishing scalar spin chirality based on the multipole cluster theory \cite{Suzuki_PRB_2021}. In this sense, the overall wave function of the CSL in 30 nm Y$_2$Ir$_2$O$_7$ must include spin configurations containing these chiral bases, such as `two-in-two-out', `three-in-one-out', and/or other linearly combined ones, which is analogous to the Pr spin configurations in the metallic CSL state of Pr$_2$Ir$_2$O$_7$ \cite{Machida_nature_2010}. 

On the other hand, theoretical calculations reveal that the spin configurations of pyrochlore iridates are located very close in energy, within several meV \cite{Wan_PRB_2011,Wang_PRL_2017}. This, in turn, implies the possibility of interconversion among the magnetic states by tuning the magnitudes of direct and indirect hopping between the Ir atoms \cite{WK_PRB_2012,Chen_PRR_2020,Goswami_prb_2017,Ladovrechis_PRB_2021}. Based on this observation, we speculate that in approaching the quasi-2D limit, truncation of the pyrochlore lattice along the (111) orientation and enhanced fluctuations rescale the relative strength of hopping and magnetic interactions, leading to a degenerate ground-state manifold. Fig.~\ref{structure}d provides one of the plausible snapshots of the CSL state in 30 nm Y$_2$Ir$_2$O$_7$, where red and blue dimers denote the distribution of fluctuating short-range spin textures. 

In summary, we unravel the abnormally ``fragile'' nature of the AIAO order and unlock a fresh region that has not been recognized on the global phase diagram of pyrochlore iridates. Moreover, the emergence of the CSL state enables new opportunities to explore other exotic quantum phases proposed for pyrochlore iridates, particularly quantum nematic \cite{Goswami_prb_2017}, axion insulator \cite{Varnava_PRB_2018}, and unconventional superconductivity \cite{Fiete_PRL_2017}. Future experiments at milli-Kelvin temperature and ultra-high magnetic field are essential to verify if the dynamic CSL state can persist or if the quantum-order-by-disorder mechanism eventually prevails and selects a static configuration.

%In addition, a cover letter needs to be written with the
%following:
%\begin{enumerate}
 %\item A 100 word or less summary indicating on scientific grounds
%why the paper should be considered for a wide-ranging journal like
%\textsl{Nature} instead of a more narrowly focussed journal.
 %\item A 100 word or less summary aimed at a non-scientific audience,
%written at the level of a national newspaper.  It may be used for
%\textsl{Nature}'s press release or other general publicity.
 %\item The cover letter should state clearly what is included as the
%submission, including number of figures, supporting manuscripts
%and any Supplementary Information (specifying number of items and
%format).
 %\item The cover letter should also state the number of
%words of text in the paper; the number of figures and parts of
%figures (for example, 4 figures, comprising 16 separate panels in
%total); a rough estimate of the desired final size of figures in
%terms of number of pages; and a full current postal address,
%telephone and fax numbers, and current e-mail address.
%\end{enumerate}

\begin{addendum}
\item [Acknowledgements.]
The authors deeply acknowledge J. Pixley, G. Fiete, X. Hu, P. Laurell, D. Vanderbilt, D. Khomskii, H. Kim for numerous insightful discussions. X.L. and J.G. acknowledge the support by the National Key R\&D Program of China (Grant No. 2022YFA1403400), the National Natural Science Foundation of China (Grant No. 12250710675, 12204521), and the Strategic Priority Research Program of the Chinese Academy of Sciences (No. XDB33000000). M.K., M.T., T.W. and J.C. acknowledge the support by the U.S. Department of Energy, Office of Science, Office of Basic Energy Sciences under award number DE-SC0022160. Y.Q. acknowledge the support by the National Natural Science Foundation of China (Grant No. 12374144). H.C. and W.H acknowledge support from the U.S. Department of Energy, Office of Science, Office of Basic Energy Sciences Early Career Research Program under Award Number DE-SC-0021305.
This research used resources of the Advanced Photon Source, a U.S. Department of Energy Office of Science User Facility operated by Argonne National Laboratory under Contract No. DE-AC02-06CH11357. The RIXS experiment and data analysis were supported in part by the U.S. Department of Energy, Office of Science, Basic Energy Sciences, Materials Science and Engineering Division. 
 
\item [Author contributions.] 
X.L. and J.C. conceived the project. X.L. and M.K. developed the sample fabrications. X.L. performed the transport measurements, and carried out the synchrotron X-ray magnetic circular dichroism experiments with assistance from YS.C. and D.H.. J.K. and P.R. performed the resonant X-ray scattering experiments. M.T., X.J. H.C. and Y.C. carried out the resonant inelastic X-ray scattering experiments with assistance from M.U. and J.K.. Theoretical analyses on magnetic structures were carried out by Y.W. and Y.Q.. The manuscript was written by X.L., Y.W. and J.C.. All authors discussed the results and commented on the manuscript.
 
\item [Additional information.] 
Supplementary information is available in the online version of this manuscript. Reprints and permissions information is available online at www.nature.com/reprints. Correspondence and requests for materials should be addressed to X.L. (xiaoran.liu@iphy.ac.cn).

\item [Competing interests.]
The authors declare no competing financial interests. 
\end{addendum}

%\subsection{Method subsection.}
%Here is a description of a specific method used.  Note that the
%subsection heading ends with a full stop (period) and that the
%command is \verb|\subsection{}| not \verb|\subsection*{}|.

%% Put the bibliography here, most people will use BiBTeX in
%% which case the environment below should be replaced with
%% the \bibliography{} command.

%% Here is the endmatter stuff: Supplementary Info, etc.
%% Use \item's to separate, default label is "Acknowledgements"

%%
%% TABLES
%%
%% If there are any tables, put them here.
%%

\clearpage

\begin{methods}

\noindent{\bf {Sample fabrication.}}\ The (111) oriented Y$_2$Ir$_2$O$_7$ thin films were fabricated on 5$\times$5 mm$^2$ (111) yttria-doped ZrO$_2$ (YSZ) substrates by pulsed laser deposition. Ir-rich phase-mixed ceramic targets (Y:Ir = 1:3) were ablated using a KrF excimer laser ($\lambda$ = 248 nm, energy density $\sim$ 5 J/cm$^{2}$) with a repetition rate of 10 Hz. The deposition was first carried out at a substrate temperature of 450 $^{\circ}$C, under 100 mTorr atmosphere composed of Ar and O$_2$ gases (partial pressure ratio, Ar:O$_2$  = 10:1). Amorphous film with the proper stoichiometry was obtained from this stage. Then the film was post-annealed inside the chamber at 950 $^{\circ}$C under 500 Torr atmosphere of pure O$_2$ for 15 - 20 min, followed by cooling to room temperature. High-quality single crystalline film was eventually achieved.\

%\noindent{\bf {Scanning transmission electron microscopy.}}\ The scanning transmission electron microscopy (STEM) measurements were carried out using a spherical aberration-corrected JEM-ARM200F, operated at 200 kV. The high-angle annular dark-field (HAADF) imaging was taken using the collection semi-angle of about 70-250 mrad.\ 

\noindent{\bf {Transport measurements.}}\ The electrical transport measurements were performed in a 9 T physical property measurement system (PPMS, Quantum Design) with the 4-point contact method. The magnetic field was applied along the [111] direction, and the current was driven along the [1$\bar{1}$0] direction. In particular, due to the extremely large coercivity, the magneto-transport at each temperature was measured in the following procedures. First, +9 T was applied at 150 K and the sample was field cooled to the desired temperature. The resistance was measured by sweeping field from +9 to -9 T. Then, sample was heated to 150 K (to quench any magnetic ordering), and cooled back to the same temperature, after which the resistance was measured again by sweeping field from -9 to +9 T. Symmetrization (anti-symmetrization) was further applied using these two branches on the longitudinal (transverse) resistance to achieve pure signal from the magnetoresistance (the Hall effect).\       

\noindent{\bf {Resonant x-ray scattering.}}\ The resonant x-ray scattering experiments were performed at beamline 6-ID-B of the Advanced Photon Source in Argonne National Laboratory using a 6-circle diffractometer. The (111) Eu$_2$Ir$_2$O$_7$ and (111) Y$_2$Ir$_2$O$_7$ films were mounted in a close-cycle displex such that the in-plane [$\bar{1}$10] direction was along the x-rays, and the [110] direction was aligned perpendicular to the scattering plane. The incident beam was linearly polarized perpendicular to the scattering plane ($\sigma$), with energy tuned around the Ir L$_3$ absorption edge. A PG (008) analyzer was used to filter the polarization of the scattered x-rays, leading to two detection channels, $\sigma$-$\sigma$ and $\sigma$-$\pi$. The signal of resonant magnetic scattering can be effectively probed in the $\sigma$-$\pi$ channel.\

\noindent{\bf {X-ray magnetic circular dichroism spectra.}}\ The resonant x-ray magnetic circular dichroism (XMCD) experiments were performed at beamline 4-ID-D of the Advanced Photon Source at Argonne National Laboratory. X-ray absorption spectra around Ir L$_3$ and L$_2$ edges were collected with left- and right-circularly polarized beams at normal incidence ($k$ // [111]) to obtain the XMCD spectra. To exclude any artifact, XMCD spectra were measured in both positive and negative external field. In case of large magnetic coercivity, the measurements were conducted in two ways for comparison. (1) XMCD was recorded after sample was +6 T field cooled to 4 K; then sample was heated to 150 K (to quench any magnetic ordering) and cooled back to 4 K with -6 T field, after which XMCD was recorded. (2) XMCD was recorded after sample was +6 T field cooled to 4 K. Then the magnetic field was directly swept to -6 T at 4 K, after which XMCD was recorded. The spectra obtained in both ways are almost identical.\

\noindent{\bf {Resonant inelastic x-ray scattering.}}\ Resonant inelastic X-ray scattering (RIXS) measurements were taken at beamline 27-ID-B of Advanced Photon Source. The incoming photon energy was tuned to the Ir $L_3$ pre-edge (11.215 keV) with 28 meV resolution (FWHM) to enhance the magnetic scattering cross-section. Measurements on thin films of several tens' nm at such a hard x-ray facility imposed several constraints on the sample geometry: (a) grazing-incident geometry to guarantee the maximal footprint of x-ray beams on the film, enhancing the overall scattering signal; (b) 2$\theta$ $\approx$ 90$^\circ$ to minimize the elastic scattering signal and to purify the magnetic signal. Furthermore, the magnetic signal was enhanced by selecting a peak which was structurally forbidden but magnetically allowed. As a result, measurements were taken around the (5 12 -1) peak.
To obtain the exact information of these peaks, the elastic line was fitted to a Voigt function with a resolution determined at $\Gamma$ point. The inelastic peaks were fitted to anti-symmetrized Lorentzian line shapes of the form:
\begin{align}
S(Q,\omega) = \frac{A}{2\pi(1-exp(\frac{\omega}{k_BT}))} \gamma (\frac{1}{(\omega-\omega_0)^2 + (\gamma/2)^2} + \frac{1}{(\omega+\omega_0)^2 + (\gamma/2)^2})
\end{align}
where A is the peak amplitutde, $\frac{1}{1-exp(\frac{\omega}{k_BT})}$ is the Bose factor, $\gamma$ is the peak width, and $\omega_0$ is the peak position.

%\noindent{\bf {Data availability.}}\ All data that support the findings of this study are available from the corresponding author on request.\\

\end{methods}

%\begin{thebibliography}{99}

%\bibitem{Balents_ARCMP_2014}
%Witczak-Krempa, W., Chen, G., Kim, Y. \& Balents, L.~Correlated quantum phenomena in the strong spin-orbit regime. {\it{Annu. Rev. Condens. Matter Phys.}} \textbf{5}, 57-82 (2014).

%\end{thebibliography}

\end{document}